\documentclass[prb,twocolumn,showpacs]{revtex4}

\usepackage{bm}
\usepackage{graphicx}

\newcommand{\be}{\begin{equation}}
\newcommand{\ee}{\end{equation}}
\newcommand{\ba}{\begin{eqnarray}}
\newcommand{\ea}{\end{eqnarray}}

\newcommand{\Dv}{D_v}

\begin{document}

\title{Phase diagram of underdoped cuprate superconductors:\\
effect of Cooper-pair phase fluctuations}

\author{C. Timm}
\email{timm@physik.fu-berlin.de}
\author{D. Manske}
\author{K. H. Bennemann}
\affiliation{Institut f\"ur Theoretische Physik, Freie Universit\"at
Berlin, Arnimallee 14, D-14195 Berlin, Germany}
\date{February 16, 2002}

\begin{abstract}
In underdoped cuprates fluctuations of the phase of the superconducting
order parameter play a role due to the small
superfluid density. We consider the effects of phase fluctuations assuming
the exchange of spin fluctuations to be the predominant pairing
interaction. Spin fluctuations are treated in the fluctuation-exchange
approximation, while phase fluctuations are included by
Berezinskii-Kosterlitz-Thouless theory. We calculate the
stiffness against phase fluctuations, $n_s(\omega)/m^\ast$, as a function
of doping, temperature, and frequency, taking its renormalization by phase
fluctuations into account. The results are compared with recent
measurements of the high-frequency conductivity. Furthermore, we obtain the
temperature $T^\ast$, where the density of states at the
Fermi energy starts to be suppressed, the temperature
$T_c^\ast$, where Cooper pairs form, and the superconducting
transition temperature $T_c$, where their phase becomes coherent. We find a
crossover from a phase-fluctuation-dominated regime with $T_c\propto n_s$
for underdoped cuprates to a BCS-like regime for overdoped materials.
\end{abstract}


\maketitle

\section{Introduction}

For about fifteen years cuprate high-temperature superconductors (HTSC's)
have stimulated significant advances in the theory of highly correlated
systems as well as in soft condensed matter theory. Nevertheless, we still
do not fully understand the various phases of these materials. Of
particular interest is the underdoped regime of hole-doped cuprates, in
which the hole density (doping) $x$ in the CuO$_2$ planes is lower than
required for the maximum superconducting transition temperature $T_c$. In
this regime the superfluid density $n_s$ decreases with decreasing doping
and is found to be proportional to $T_c$.\cite{Uemura} Above $T_c$, one
finds a strong suppression of the electronic density of states close to the
Fermi energy, {\it i.e.}, a {\it pseudogap}, which appears to have the same
symmetry as the superconducting gap.\cite{TK} Furthermore, there may be
fluctuating charge and spin modulations (stripes).\cite{strip}

It has been recognized early on that the small superfluid density $n_s$
leads to a reduced stiffness against fluctuations of the phase of the
superconducting order parameter.\cite{DI,Cha,EK95} Phase fluctuations are
additionally enhanced because they are canonically conjugate to charge
density fluctuations, which are believed to be
suppressed.\cite{DI,EK95} Furthermore, the cuprates
consist of weakly coupled two-dimensional (2D) CuO$_2$ planes so that
fluctuations are enhanced by the reduced dimensionality. Phase fluctuations
might destroy the long-range superconducting order, although there is still
a condensate of preformed Cooper pairs. In conventional, bulk
superconductors this mechanism is not relevant, since the large superfluid
density leads to a typical energy scale of phase fluctuations much higher
than the superconducting energy gap $\Delta$, which governs the thermal
breaking of Cooper pairs. Thus in conventional superconductors the
transition is due to the destruction of the Cooper pairs and $T_c$ is
proportional to $\Delta$.\cite{Schbook} On the other hand, the observation
that $T_c\propto n_s$ in underdoped cuprates\cite{Uemura} indicates that
the phase fluctuations drive the transition in this regime. The Cooper
pairs only break up at a crossover around $T_c^\ast>T_c$. If the feedback
of phase fluctuations on the local formation of Cooper pairs is small,
$T_c^\ast$ is approximately given by the transition temperature one would
obtain without phase fluctuations. Between $T_c$ and $T_c^\ast$ Cooper
pairs exist but the order parameter is not phase
coherent.\cite{DI,Cha,EK95,FS,SGB} Recent thermal-expansion experiments
strongly support this picture.\cite{Meingast} However, there is no close
relation between our $T_c^\ast$ and the mean-field transition temperature
of Ref.~\onlinecite{Meingast}, which is determined by extrapolation from
the low-$T$ behavior of the expansivity.

There is a third temperature scale $T^\ast$ with $T^\ast>T_c^\ast$, below
which a pseudogap starts to open up as seen in nuclear magnetic resonance,
tunneling, and transport experiments.\cite{Ta1,Ta2,Ta3,BT,TS} It seems
unlikely that the pseudogap at these temperatures is due to local
superconductivity. Rather, it is thought to be caused by spin
fluctuations\cite{SGB} or the onset of stripe
inhomogeneities.\cite{BE,Mark} Recent experiments on the Hall effect in
GdBa$_2$Cu$_3$O$_{7-\delta}$ films\cite{Matthey} also support the existence
of {\it two\/} crossover temperatures $T_c^\ast$ and $T^\ast$. In this work
we are mostly concerned with the strong pseudogap regime $T_c<T<T_c^\ast$.

Due to the layered structure of the cuprates, they behave like the 2D {\it
XY\/} model except in a narrow critical range around $T_c$, where they show
three-dimensional (3D) {\it XY\/} critical behavior.\cite{Feig79,Blt} The
standard theory for the 2D {\it XY\/} model, the
Berezinskii-Kosterlitz-Thouless (BKT) renormalization group
theory,\cite{BKT,BDHT79,Minn,KR} should thus describe these materials
outside of the narrow critical
range.\cite{Pier1,Frie,CTlay,CTflux,Pier2,FATZ} Also, recent transport
measurements for a gate-doped cuprate\cite{Schoen} with only a single
superconducting CuO$_2$ plane show essentially the same doping dependence
of $T_c$ as found for bulk materials. BKT theory predicts a transition at a
temperature $T_c<T_c^\ast$, due to the unbindung of fluctuating
vortex-antivortex pairs in the superconducting order parameter. Gaussian
phase fluctuations are less important, since they do not shift
$T_c$.\cite{rem.gauss} In addition, the coupling of the phase to the
electromagnetic field causes them to be gapped at the plasma
frequency,\cite{And} whereas the BKT picture of vortex unbinding remains
basically unchanged.\cite{BC}

In the early days of HTSC's, BKT theory was envoked to interpret a number
of experiments on bulk samples.\cite{Stam,Arte,Mart,Yeh,Frel,Prad}
Recently, two experiments have lent strong additional support to the BKT
description: First, Corson {\it et al.}\cite{Cors} have measured the
complex conductivity of underdoped Bi$_2$Sr$_2$CaCu$_2$O$_{8+\delta}$ and
extracted the frequency-dependent phase stiffness from the data. The
authors interpret their data in terms of dynamical vortex-pair
fluctuations\cite{AHNS,Minn} and conclude that vortices---and thus a local
superconducting condensate---exist up to at least $100\,{\rm K}$. We
discuss this assertion in Sec.~\ref{sec.3}. Second, Xu {\it et
al.}\cite{Xu} have found signs of vortices at temperatures much higher than
$T_c$ in underdoped La$_{2-x}$Sr$_x$CuO$_4$ in measurements of the Nernst
effect. A recent reanalysis of the data\cite{Wang} yields an onset
temperature of vortex effects of $40\,{\rm K}$ for an extremely underdoped
sample ($x=0.05$) and of $90\,{\rm K}$ for $x=0.07$.

So far, we have not said anything about the superconducting pairing
mechanism. There is increasing evidence that pairing is mainly due to the
exchange of spin fluctuations. The conserving fluctuation-exchange (FLEX)
approximation\cite{Bickers,Pao,Monthoux,DahmTewordt,Bgroup,SGB} based on
this mechanism describes optimally doped and overdoped cuprates rather
well. In particular, the correct doping dependence and order of magnitude
of $T_c$ are obtained in this regime. On the other hand, the FLEX
approximation does not include phase fluctuations and we believe this to be
the main reason why it fails to predict the downturn of $T_c$ in the
underdoped regime. Instead, $T_c$ is found to approximatetly saturate for
small doping $x$. However, the FLEX approximation is able to reproduce two
other salient features of underdoped cuprates, namely the decrease of $n_s$
and the opening of a weak pseudogap at $T^\ast$, as we show below.

This encourages us to apply the following description. We employ the FLEX
approximation to obtain the dynamical phase stiffness $n_s(\omega)/m^\ast$,
where $n_s(\omega)$ is the generalization of the superfluid density for
finite frequencies. The static density $n_s(0)$ starts to deviate from zero
at the temperature where Cooper pairs start to form and which we identify
with $T_c^\ast$. Then, phase fluctuations are incorporated by using the
phase stiffness from FLEX as the input for BKT theory, which leads to a
renormalized $n_s^R<n_s$ and predicts a reduced $T_c$. Then, we consider
the dynamical case $\omega>0$ and use dynamical BKT theory\cite{AHNS,Minn}
to find the renormalized phase stiffness $n_s^R(\omega)/m^\ast$ and compare
the results with experiments.\cite{Cors}

\section{Static case}
\label{sec.2}

Transport measurements for a gate-doped cuprate\cite{Schoen} show that the
superconducting properties are determined by a single CuO$_2$ plane. The
simplest model believed to contain the relevant strong correlations is the
2D one-band Hubbard model.\cite{AndHT} We here start from the Hamiltonian
\begin{equation}
H = - \sum_{\langle ij \rangle \, \sigma}
t_{ij}\left( c_{i\sigma}^\dagger c_{j\sigma} +
c_{j\sigma}^\dagger c_{i\sigma}\right)
+ U\, \sum_i n_{i\uparrow}n_{i\downarrow} .
\label{eq:hubbard}
\end{equation}
Here, $c_{i\sigma}^\dagger$ creates an electron with spin $\sigma$
on site $i$, $U$ denotes the on-site Coulomb interaction,
and $t_{ij}$ is the hopping integral.
Within a conserving approximation, the one-electron self-energy
is given by the functional derivative of a generating functional
$\Phi$, which is related to the free energy, with respect
to the dressed one-electron Green function ${\cal G}$,
$\Sigma=\delta\Phi[H]/\delta{\cal G}$.\cite{BK}
On the other hand, the dressed Green function is given by the usual Dyson
equation ${\cal G}^{-1} = {\cal G}_0^{-1} - \Sigma$ in terms of the
unperturbed Green function ${\cal G}_0$ of the kinetic part of $H$ alone.
These equations determine the dressed Green function.\cite{BK}

The $T$-matrix\cite{Tewordt} or FLEX
approximation\cite{Bickers,Pao,Monthoux,DahmTewordt,Bgroup,SGB} is
distinguished by the choice of a particular infinite subset of ladder and
bubble diagrams for the generating functional $\Phi$. The dressed Green
functions are used to calculate the charge and spin susceptibilities. From
these a Berk-Schrieffer-type\cite{BerkSchrieffer} pairing interaction is
contructed, describing the exchange of charge and spin fluctuations. In a
purely electronic pairing theory a self-consistent description is required
because the electrons do not only form Cooper pairs but also mediate the
pairing interaction. The quasiparticle self-energy components $X_{\nu}$
($\nu= 0$, $3$, $1$) with respect to the Pauli matrices $\tau_{\nu}$ in the
Nambu representation,\cite{Nambu,Schbook} {\it i.e.}, $X_0=\omega(1-Z)$
(renormalization), $X_3=\xi$ (energy shift), and $X_1=\phi$ (gap parameter)
are given by
\begin{eqnarray}
X_{\nu}({\bf k},\omega) & = & \frac{1}{N}
\sum_{{\bf k'}} \int_0^{\infty} \!\!\!
d\Omega \, \left[ P_s({\bf k}\!-\!{\bf k'},\Omega) \pm
P_c({\bf k}\!-\!{\bf k'},\Omega)\right] \nonumber \\
& \times &
\int_{-\infty}^{\infty} d\omega' \, I(\omega,\Omega,\omega')\,
A_{\nu}({\bf k'},\omega') .
\label{eq:selfenergy}
\end{eqnarray}
Here, the plus sign holds for $X_0$ and $X_3$ and the minus sign
for $X_1$. The kernel $I$ and the spectral functions $A_{\nu}$
are given by
\begin{eqnarray}
I(\omega,\Omega,\omega') & = &
  \frac{f(-\omega') + b(\Omega)}{\omega + i\delta - \Omega - \omega'} +
  \frac{f(\omega') + b(\Omega)}{\omega + i\delta + \Omega - \omega'} , \\
A_{\nu}({\bf k},\omega) & = & -\frac{1}{\pi}\:
  \mbox{Im}\, \frac{a_{\nu}({\bf k},\omega)}{D({\bf k},\omega)} ,
\end{eqnarray}
where
$a_0 = \omega Z$,
$a_3 = \epsilon_{\bf k} + \xi$,
$a_1 = \phi$, and
\begin{equation}
D = (\omega Z)^2 - \left[ \epsilon_{\bf k} +
\xi\right]^2 - \phi^2 .
\end{equation}
Here, $f$ and $b$ are the Fermi and Bose distribution
function, respectively.
We use the bare tight-binding dispersion relation for lattice constant
$a=b=1$,
\be
\epsilon_{\bf k}=2t\,(2 - \cos k_x - \cos k_y - \mu) .
\ee
The band filling $n=1/N \sum_{\bf k}n_{\bf k}$ is
determined with the help of the ${\bf k}$-dependent occupation
number $n_{\bf k}=2\int_{-\infty}^{\infty}d\omega\, f(\omega)\,
N({\bf k},\omega)$ which is calculated self-consistently. $n=1$
corresponds to half filling. The interactions due to
spin and charge fluctuations are given by
$P_s=(2\pi)^{-1}U^2\,\mbox{Im}\,(3\chi_s-\chi_{s0})$ with
$\chi_s=\chi_{s0}\,(1-U\chi_{s0})^{-1}$ and
$P_c=(2\pi)^{-1}U^2\,\mbox{Im}\,(3\chi_c-\chi_{c0})$ with
$\chi_c=\chi_{c0}\,(1+U\chi_{c0})^{-1}$. In terms of spectral
functions one has
\begin{eqnarray}
\lefteqn{
\mbox{Im}\,\chi_{s0,c0}({\bf q},\omega)
  = \frac{\pi}{N} \int_{-\infty}^{\infty} \!\!\!
  d\omega' \, \left[ f(\omega') - f(\omega' + \omega)\right] }
  \nonumber \\
& & {}\times \sum_{{\bf k}} \big[ N({\bf k} + {\bf q},\omega' + \omega)\,
  N({\bf k},\omega') \nonumber \\
& & \quad{}\pm
  A_1({\bf k} + {\bf q},\omega' + \omega)\, A_1({\bf k},\omega')
  \big] . \qquad\qquad
\label{eq:barechi}
\end{eqnarray}
Here, $N({\bf k},\omega)= A_0({\bf k},\omega) + A_3({\bf k},\omega)$,
and the real parts are calculated with the help of the
Kramers-Kronig relation. The substracted terms in $P_s$ and $P_c$
remove double counting that occurs in second order.
The spin fluctuations are found to dominate the pairing interaction.
The numerical calculations are performed on a square lattice
with $256\times 256$ points in the Brillouin zone
and with $200$ points on the real $\omega$ axis up to $16\,t$ with
an almost logarithmic mesh. The full momentum and frequency dependence
of the quantities is kept.
The convolutions in ${\bf k}$ space are carried out using
fast Fourier transformation.
The superconducting state is found to have $d_{x^2-y^2}$-wave symmetry.
$T_c^\ast$ is determined from the linearized gap equation.

A field-theoretical derivation of the effective action of phase
fluctuations\cite{KD,SBL,rem.fieldth,FS} shows that the phase stiffness for
frequency $\omega=0$ is given by the 3D static superfluid density divided by
the effective mass, $n_s(x,T)/m^\ast$. This quantity is given by
\begin{equation}
\frac{n_{s}}{m^\ast} = \frac{2}{\pi e^2}\, (I_N - I_S)
\label{eq:snss}
\end{equation}
with
\begin{equation}
I_{N,S} = \int_0^{\infty} \!\!\! d\omega\: \sigma_1^{N,S}(\omega) ,
\end{equation}
where $\sigma_1^N$ ($\sigma_1^S$) is the real part of the conductivity in
the normal (superconducting) state. Here we utilize the $f$-sum rule
$\int_0^{\infty} d\omega\,\sigma_1(\omega)=\pi e^2n/2m^\ast$
where $n$ is the 3D electron density. The interpretation of
Eq.~(\ref{eq:snss}) is that the
spectral weight missing from the quasi-particle background in
$\sigma(\omega)$ for $T<T_c^\ast$
must be in the superconducting delta-function peak.

$\sigma(\omega)$ is calculated in the normal
and superconducting states using the Kubo formula\cite{Mahan,Wermbter}
\begin{eqnarray}
\sigma(\omega) & = & \frac{2 e^2}{\hbar c}\,
  \frac{\pi}{\omega} \int_{-\infty}^{\infty} \!\!\! d\omega'
\left[f(\omega') - f(\omega'+\omega)\right]
\nonumber\\
& & {}\times
\frac{1}{N} \sum_{\bf k} (v_{{\bf k},x}^2 + v_{{\bf k},y}^2)\,
  \big[ N({\bf k},\omega'+\omega)\, N({\bf k},\omega')
  \nonumber \\
& & \quad{} + A_1({\bf k},\omega'+\omega)\,
  A_1({\bf k},\omega') \big] ,
\end{eqnarray}
where $v_{{\bf k},i}=\partial\epsilon_{\bf k}/\partial k_i$ are the
band velocities within the $\mbox{CuO}_2$ plane and
$c$ is the $c$-axis lattice constant.
Vertex corrections are neglected.

\begin{figure}
\includegraphics[width=3.20in]{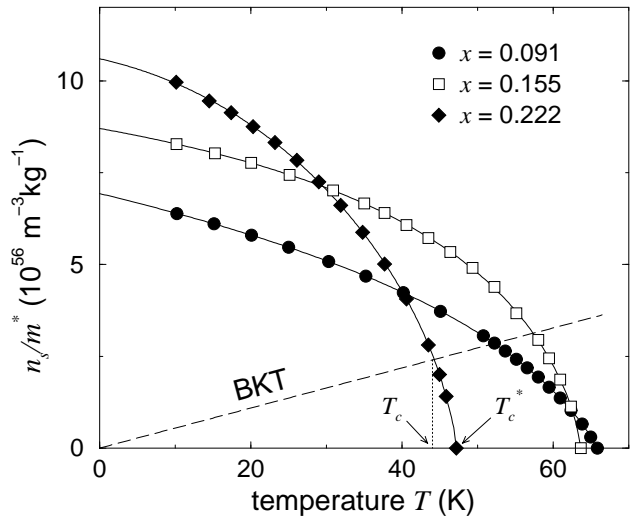}
\caption{\label{fig.ns0}Static superfluid density as a function of
temperature for three values of the doping $x$ (symbols). The solid curves
are fits of power laws with logarithmic corrections as explained in the
text. The intersection of $n_s(T)/m^\ast$ with the dashed line represents a
simplified criterion for the BKT transition temperature $T_c$.}
\end{figure}

The superfluid density (phase stiffness) $n_s/m^\ast$ obtained in this way
is shown in Fig.~\ref{fig.ns0} for the three doping values $x=0.091$
(underdoped), $x=0.155$ (approximately optimally doped), and $x=0.222$
(overdoped). The figure also shows fits to the data at given doping level,
where we assume the form $\ln n_s(T)/m^\ast \cong a_0 + a_1 \ln(T_c^\ast-T)
+ a_2 \ln^2(T_c^\ast-T) +\ldots$, {\it i.e.}, a power-law dependence
close to $T_c^\ast$ with logarithmic corrections. We use the
fits to extrapolate to $T=0$. The results show that $T_c^\ast$ depends on
$x$ only weakly in the underdoped regime but decreases rapidly in the
overdoped. We come back to this below. Furthermore, $n_s/m^\ast$ increases
much more slowly below $T_c^\ast$ in the underdoped regime and extrapolates
to a smaller value at $T=0$.

\begin{figure}
\includegraphics[width=3.20in]{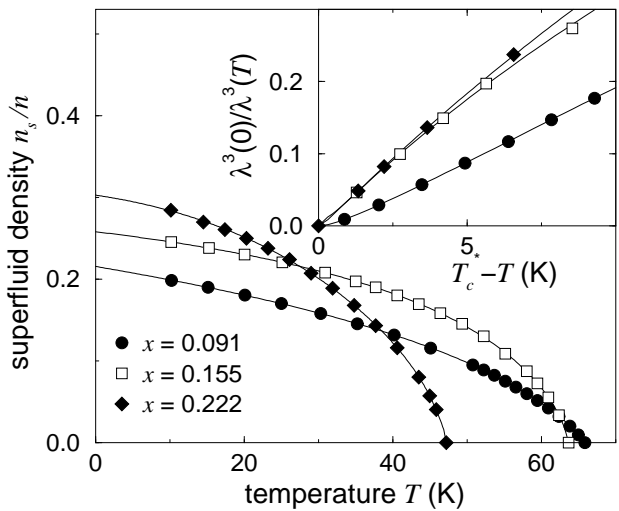}
\caption{\label{fig.nsn}Ratio of superfluid density to total hole density
for the same doping values $x$ as in Fig.~\protect\ref{fig.ns0}. The inset
shows $\lambda^3(T\!=\!0)/\lambda^3(T) = n_s^{3/2}(T)/n_s^{3/2}(T\!=\!0)$,
where $\lambda$ is the London penetration depth, as a function of
$(T_c^\ast-T)$.}
\end{figure}

We have also calculated $n_s$ in units of the total hole density $n$, shown
in Fig.~\ref{fig.nsn}, finding that $n_s/n$ is significanly reduced below
unity, in agreement with experiments but in contradiction to BCS theory.
The reduction is strongest for the underdoped case. Our results show that
spin fluctuations can explain most of the observed reduction of $n_s$. Also
note that $n_s$ is linear in temperature for $T\to 0$ because of the nodes
in the gap. The inset in Fig.~\ref{fig.nsn} shows
$\lambda^3(0)/\lambda^3(T)$, where the penetration depth is\cite{Schbook}
$\lambda\propto n_s^{-1/2}$, as a function of $T_c^\ast-T$. The FLEX
approximation yields $\lambda^3(0)/\lambda^3(T)\propto T_c^\ast-T$. The
same power law has been found experimentally by Kamal {\it et
al}.\cite{Kamal} It has been attributed to critical fluctuations starting
about $10\,{\rm K}$ below the transition temperature,\cite{Kamal} since it
coincides with the critical exponent expected for the 3D {\it XY\/} model.
We here obtain {\it the same\/} power law from the FLEX approximation,
which is purely 2D and does not contain critical fluctuations. Instead this
rapid increase of $n_s\propto 1/\lambda^2$ below $T_c^\ast$ is due to the
selfconsistency, which leads to a more rapid opening of the gap than in BCS
theory. We thus conclude that, while critical 3D {\it XY\/} fluctuations
are expected in a narrow temperature range,\cite{Feig79,Blt} they are
not the origin of the observed power law on the scale of $10\,{\mathrm K}$.

Now we turn to the renormalization of $n_s$ due to phase (vortex)
fluctuations. The
BKT theory describes the unbinding of thermally created pancake
vortex-antivortex pairs.\cite{BKT,Minn} The relevant parameters
are the dimensionless stiffness $K$ and the core energy $E_c$
of vortices. The stiffness is related to $n_s$ by\cite{rem.EK}
\be
K(T) = \beta\hbar^2\,\frac{n_s(T)}{m^\ast}\,\frac{d}{4} ,
\label{2.Kofns}
\ee
where $\beta$ is the inverse temperature and
$d$ is the average spacing between CuO$_2$ layers. Since we use a
2D model to describe double-layer cuprates, we set $d$ to half
the height of the unit cell of the typical representative
YBa$_2$Cu$_3$O$_{6+y}$. The stiffness
$K$ is also a measure of the strength of the vortex-antivortex interaction
$V = 2\pi k_B T K \,\ln (r/r_0)$. Here, $r_0$ is the minimum pair size,
{\it i.e.}, twice the vortex core radius, which is of the order of
the in-plane Ginzburg-Landau coherence length $\xi_{ab}$. For the core
energy we use an approximate result by Blatter {\it et al.},\cite{Blt}
$E_c = \pi k_B T\, K\, \ln\kappa$,
where $\kappa$ is the Ginzburg parameter.
Starting from the smallest vortex-antivortex pairs of size $r_0$,
the pairs are integrated out and their effect is incorporated by an
approximate renormalization of $K$ and the fugacity\cite{rem.y}
$y = e^{-\beta E_c}$.
This leads to the Kosterlitz recursion relations
\ba
\frac{d y}{d l} & = & (2-\pi K)\, y ,
\label{2.Kost1} \\
\frac{d K}{d l} & = & -4\pi^3 y^2 K^2 ,
\label{2.Kost2}
\ea
where $l=\ln (r/r_0)$ is a logarithmic length scale. For $T>T_c$, $K$
goes to zero for $l\to\infty$, so that the interaction is screened
at large distances and the largest vortex-antivortex pairs unbind.
The unbound vortices destroy the superconducting order and the Mei\ss{}ner
effect and lead to
dissipation.\cite{Mooij} For $T<T_c$, $K$ approaches a finite value
$K_R\equiv \lim_{l\to\infty} K$ and $y$
vanishes in the limit $l\to\infty$ so that there are exponentially few
large pairs and they still feel the logarithmic interaction. Bound pairs
reduce $K$ and thus $n_s$, but do not destroy superconductivity. At
$T_c$, $K_R$ jumps from a universal value of $2/\pi$ to zero.
The values of $T_c$ shown below are obtained by numerically integrating
Eqs.~(\ref{2.Kost1}) and (\ref{2.Kost2}) with $n_s$ taken from an
interpolation between the points in Fig.~\ref{fig.ns0}. It turns out that the
renormalization of $K$ for $T<T_c$ is very small so that one obtains
$T_c$ from the simple criterion
\be
K(T_c) = \frac{2}{\pi} \quad\mbox{or}\quad
\frac{n_s(T_c)}{m^\ast} = \frac{2}{\pi}\, \frac{4 k_B T_c}{\hbar^2 d}
\label{2.BKTcond}
\ee
for the {\it unrenormalized\/} stiffness with an error of less than $1\,\%$.
Eq.~(\ref{2.BKTcond}) is satisfied at the intersection of the $n_s(T)/m^\ast$
curves with the dashed straight line in Fig.~\ref{fig.ns0}.

\begin{figure}[ht]
\includegraphics[width=3.20in]{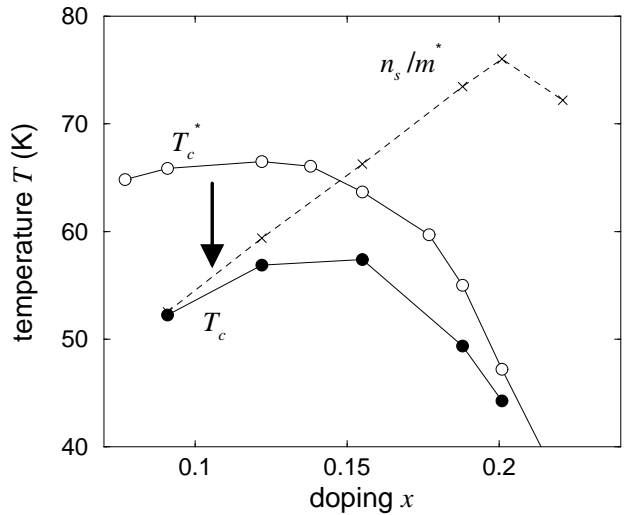}
\caption{\label{fig.T}Temperature scales of the cuprates as functions of
doping $x$. $T_c$ (solid circles) is the transition temperature obtained
from the FLEX approximation with phase fluctuations included by means of
BKT theory. At $T_c^\ast$ (open circles) Cooper pairs start to form
locally; this temperature is given by the transition temperature obtained
from the FLEX approximation with spin fluctuations alone.
The crosses show the superfluid density (phase stiffness)
$n_s(T=0)/m^\ast$ for comparison. This curve has been scaled so that it
agrees with $T_c$ in the underdoped regime.}
\end{figure}

From BKT theory we obtain two important quantities: the transition
temperature $T_c$ and the renormalized stiffness $K_R$, which determines the
renormalized superfluid density (phase stiffness)
\be
\frac{n_s^R}{m^\ast} = \frac{4}{\beta\hbar^2 d}\, K_R .
\ee
In Fig.~\ref{fig.T} we plot the transition temperature $T_c$ and the
temperature $T_c^\ast$ where Cooper-pairs form. For decreasing doping $x$,
$T_c^\ast$ becomes nearly constant and even decreases for the lowest doping
level, consistent with the strong decrease of the onset temperature of
vortex effects at even lower doping found by Xu {\it et al.}\cite{Xu,Wang}
We have also calculated der superconducting gap $\Delta_0$
ex\-tra\-po\-la\-ted to $T=0$ (not shown). $\Delta_0$ is here defined as
half the peak-to-peak separation in the density of states. We find
approximately $\Delta_0 \propto T_c^\ast$.

Phase fluctuations lead to a downturn of $T_c$ in the underdoped regime.
However, this reduction is not as large as experimentally
observed and our value $x\approx 0.14$ for the optimal doping is accordingly
smaller than the experimental one of $x\approx 0.16$.\cite{Tallon}
We suggest that one origin of this discrepancy is the neglect of the
feedback of phase fluctuations on the electronic properties.

Figure \ref{fig.T} also shows the superfluid density $n_s(0)/m^\ast$
extrapolated to $T=0$, scaled such that it approaches $T_c$ in the
underdoped regime. The density increases approximately linearly with doping
except for the most overdoped point, where it turns down again. This
behavior agrees well with angle-resolved photoemission (ARPES)
results of Feng {\it et al}.\cite{Feng} and with recent $\mu$SR experiments
of Bernhard {\it et al.}\cite{Bern} In Ref.~\onlinecite{Bern} a maximum in
$n_s$ at a unique doping value of $x_{\mathrm{max}}\approx 0.19$ is found
for various cuprates, while we obtain $x_{\mathrm{max}}\approx 0.20$. Our
results are consistent with the Uemura scaling\cite{Uemura} $T_c \propto
n_s(0)$ in the heavily underdoped regime and with the BCS-like behavior
$T_c \approx T_c^\ast \propto \Delta_0$ in the overdoped limit. $T_c$
interpolates smoothly between the extreme cases. We find $T_c<T_c^\ast$
even for high doping, since $n_s(T)$ and $K(T)$ continuously go to zero at
$T_c^\ast$ so that Eq.~(\ref{2.BKTcond}) is only satisfied at a temperature
$T_c<T_c^\ast$. The results for the overdoped case may be changed if
amplitude fluctuations of the order parameter and their mixing with phase
fluctuations\cite{OGZB} are taken into account. Amplitude fluctuations are
governed by $\Delta$, which becomes smaller than the energy scale of phase
fluctuations in the overdoped regime.

The situation is complicated by the Josephson coupling between CuO$_2$
layers. This coupling leads to the appearance of Josephson vortex lines
connecting the pancake vortices between the layers.\cite{Blt} They induce a
{\it linear\/} component in the vortex-antivortex interaction. This
contribution becomes relevant at separations larger than
$\Lambda=d/\epsilon$, where $\epsilon<1$ is the anisotropy
parameter.\cite{Blt} $\Lambda$ acts as a cutoff for the Kosterlitz
recursion relations and eventually leads to an increase of $T_c$ relative
to the BKT result $T_c^{\mathrm{BKT}}$ and to the breakdown of 2D theory
close to the transition.\cite{Blt,Pier1,Frie,CTlay} The experiments of
Corson {\it et al.}\cite{Cors} also show that the BKT temperature
$T_c^{\mathrm{BKT}}$ extracted from the data is significantly smaller than
the experimental $T_c$. Thus $T_c$ as calculated here is a lower bound of
the true transition temperature.

The feedback of phase fluctuations on the electrons is not included in our
approach. We expect the phase fluctuations in this regime to lead to pair
breaking.\cite{FS} However, simulations of the {\it XY\/} model suggest
that this feedback is rather weak.\cite{Monien} Neglecting the feedback,
the electronic spectral function shows the unrenormalized superconducting
gap for $T_c<T<T_c^\ast$. Since there is no superconducting order in this
regime, we identify this gap with the (strong) pseudogap, which thus is
automatically $d_{x^2-y^2}$-wave-like and of the same magnitude as the
superconducting gap for $T<T_c$. Thus in this picture the pseudogap is due
to local Cooper pair formation in the absence of long-range phase
coherence. Pair breaking due to phase fluctuations should partly fill in
this gap.

\begin{figure}
\includegraphics[width=3.20in]{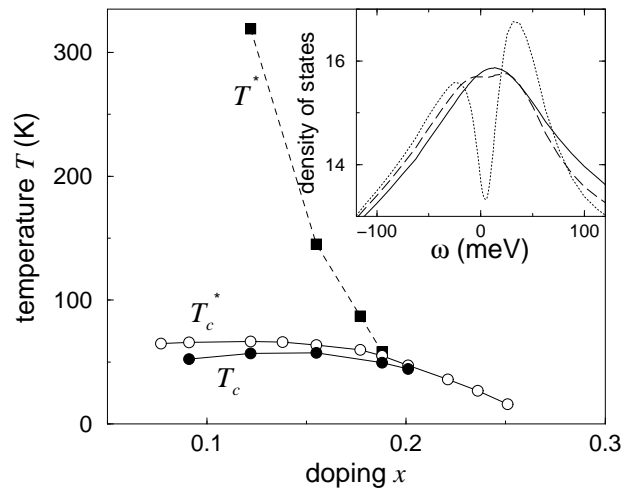}
\caption{\label{fig.T2}Temperature $T^\ast$ at which a small suppression of
the density of states at the Fermi energy (weak pseudogap) appears. The
temperatures $T_c^\ast$ and $T_c$ from Fig.~\protect\ref{fig.T} are also
shown. The inset shows the suppression of the density of states
(in arbitrary units) for $x=0.155$ and
$T=4.5\,T_c^\ast$ (solid line), $T=2.3\,T_c^\ast\approx T^\ast$ (dashed
line), and $T=1.01\,T_c^\ast$ (dotted line).}
\end{figure}

Figure \ref{fig.T2} shows $T_c$, $T_c^\ast$, and $T^\ast$ on a
different temperature scale. $T^\ast$ is the highest temperature where a
weak pseudogap is obtained from FLEX, {\it i.e.}, where the density of
states at the Fermi energy starts to be suppressed. The inset shows this
suppression for $x=0.155$. The temperature $T^\ast$ becomes much larger
than $T_c$ in the underdoped regime, in agreement with experiments.\cite{TS}

\begin{figure}
\includegraphics[width=3.20in]{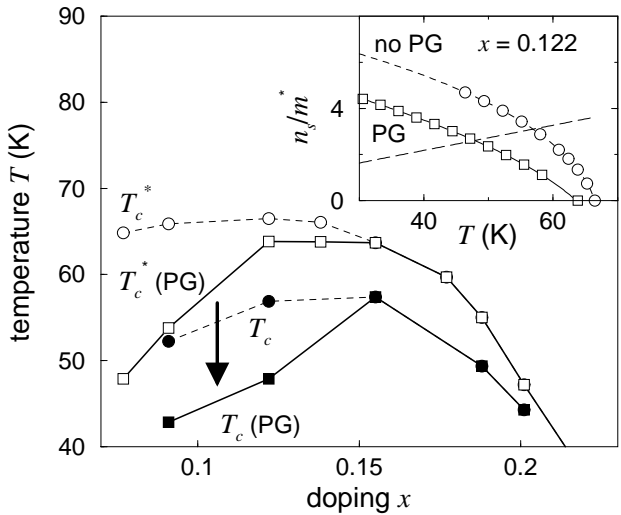}
\caption{\label{fig.TPG}Transition temperatures in the presence of a
normal-state pseudogap. The open squares show the transition
temperature $T_c^\ast$ obtained from the FLEX approximation with a $d$-wave
pseudogap in the normal-state dispersion. The amplitude of the pseudogap is
taken from experiments.\protect\cite{PG1,PG2} The open circles show the
corresponding values without a pseudogap, see Fig.~\protect\ref{fig.T}. The
solid squares denote $T_c$ in the presence of the pseudogap and with phase
fluctuations included, assuming the two effects to be independent. The
solid circles show the corresponding results without pseudogap. The inset
gives the phase stiffness $n_s/m^\ast$ for the doping $x=0.122$ with (lower
curve) and without (upper curve) the pseudogap. Intersections with the
dashed line give the simple criterion (\protect\ref{2.BKTcond}) for $T_c$.
One clearly sees that a normal-state
pseudogap increases the effect of phase fluctuations
due to the slow increase of $n_s/m^\ast$ below $T_c^\ast$.}
\end{figure}

To conclude this section, we discuss the effect of a normal-state pseudogap
due to a mechanism other than incoherent Cooper pairing. Let us assume a
suppression of the density of states close to the Fermi surface in the
normal state, {\it e.g.}, due to the formation of a charge-density
wave.\cite{DMT} This decreases the number of holes available for pairing
and should thus reduce $T_c$. To check this, we have performed FLEX
calculations with a pseudogap of the form $\Delta_{\bf k}=\Delta_0\,(\cos
k_x-\cos k_y)$ included in the normal-state dispersion. The
doping-dependent amplitude $\Delta_0$ is chosen in accordance with ARPES
experiments by Marshall {\it et al.}\cite{PG1} and by Ding {\it et
al}.\cite{PG2} The results are shown by the open squares in
Fig.~\ref{fig.TPG}. The curve merges with the $T_c^\ast$ curve without
pseudogap (open circles) at $x=0.155$, since here the pseudogap is
experimentally found to vanish.\cite{PG1,PG2} It is apparent that
$T_c^\ast$ is indeed strongly reduced in the underdoped regime. Thus this
density-of-states effect is a possible alternative explanation for the
observed downturn of $T_c$.

Next, we consider phase fluctuations in the presence of a normal-state
pseudogap. The $T_c$ values naively obtained from BKT theory for this case
are shown in Fig.~\ref{fig.TPG} as the solid squares. Phase fluctuations
reduce $T_c$ even more, in particular for $x=0.122$. This is due to the
fact that the phase stiffness $n_s/m^\ast$ increases much more slowly below
$T_c^\ast$ in the presence of a pseudogap, as shown in the inset of
Fig.~\ref{fig.TPG}, even if $T_c^\ast$ is only slightly reduced. The small
stiffness makes phase fluctuations more effective.  However, in this
picture the reduction of $T_c$ is probably overestimated: Above, we have
explained the pseudogap as resulting from incoherent Cooper pairing. This
contribution to the pseudogap must {\it not\/} be incorporated into the
normal-state dispersion to avoid double counting. This would increase the
result for $T_c$. It is clearly important to develop a theory that
incorporates phase fluctuations, spin fluctuations, and possibly the
charge-density wave on the same microscopic level. However, the inclusion
of vortex fluctuations in a FLEX-type theory on equal footing with spin
fluctuations would be a formidable task.\cite{FS}

\section{Dynamical case}
\label{sec.3}

In this section, we calculate the {\it dynamical\/} phase stiffness, which
is the quantity obtained by Corson {\it et al}.\cite{Cors} We first note
that the superfluid density can also be obtained from the imaginary part of
the conductivity,
\be
\frac{n_s}{m^\ast} = \frac{1}{e^2}\, \lim_{\omega\to 0} \omega\,
  \sigma_2^S(\omega) ,
\label{3.nsm0}
\ee
as can be shown with the help of Kramers-Kronig relations.
We have recalculated $n_s/m^\ast$ in this way and find
identical results compared to Eq.~(\ref{eq:snss}).

The phase stiffness has also been obtained at nonzero frequencies using
field-theoretical methods.\cite{KD,SBL,rem.fieldth,FS}
For small wave vector ${\bf q}\to 0$,
\be
\frac{n_s(\omega)}{m^\ast} = \frac{1}{e^2}\, \omega\, \sigma_2^S(\omega) .
\ee
The imaginary part $\sigma_2^S(\omega)$ of the dynamical conductivity is
obtained from the FLEX approximation for the dynamical current-current
correlation function using the Kubo formula.\cite{Mahan} For $\omega>0$ one
should not interpret $n_s(\omega)$ as a density. Note also that
$n_s^{-1/2}(\omega)$ is no longer proportional to the penetration depth of
a magnetic field---for $\omega>0$ there is also a contribution from the
{\it real\/} part of the conductivity, {\it i.e.}, the normal skin effect.

\begin{figure}[ht]
\includegraphics[width=3.20in]{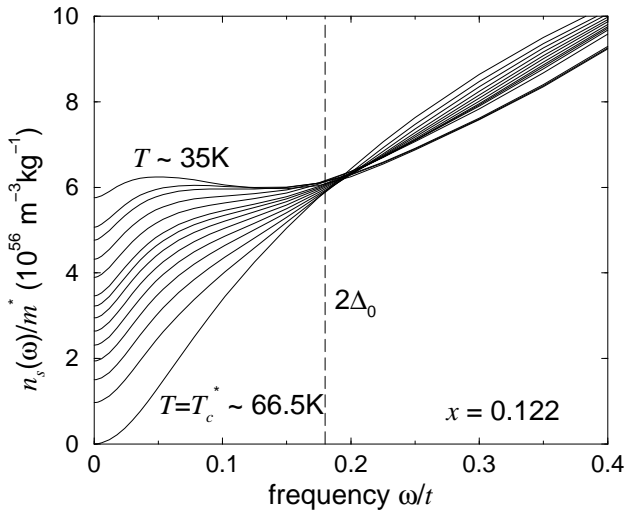}
\caption{\label{fig.nsom}Frequency-dependent phase stiffness
$n_s(\omega)/m^\ast$ for doping $x=0.122$ (underdoped) and temperatures
$k_BT/t=$ 0.012, 0.015, 0.016, 0.017, 0.018, 0.019, 0.0195, 0.02, 0.0205,
0.021, 0.0215, 0.022, 0.0225, 0.023 [with decreasing $n_s(0)/m^\ast$].
$t=250\,{\rm meV}$ is the hopping integral. The frequency is given in units
of $t$ ($\hbar=1$). At $T_c^\ast \approx 0.023\,t/k_B=66.5\,{\rm K}$ Cooper
pairs start to form. Below $T_c^\ast$ there is a marked transfer of weight
from energies above $2\Delta_0$ to energies below, where $\Delta_0$ is the
maximum gap at low temperatures as obtained from the FLEX approximation.}
\end{figure}

The resulting phase stiffness $n_s(\omega)/m^\ast$ is shown in
Fig.~\ref{fig.nsom} for $x=0.122$ (underdoped) at various temperatures. At
higher doping the results (not shown) are similar, only the typical
frequency scale, which turns out to be the low-temperature superconducting
gap $\Delta_0$, is reduced. We find a finite phase stiffness at $\omega>0$
even for $T\ge T_c^\ast$. At first glance this is surprising, since the
phase is not well-defined for $\Delta=0$. Indeed, using a Ward identity one
can show that the Gaussian part of the phase action vanishes for $T\ge
T_c^\ast$.\cite{CTWard} However, the phase action contains a contribution
from the time derivative of the phase besides the stiffness term. While the
total action vanishes, each term on its own does not. Thus the stiffness is
finite but has no physical significance for $T\ge T_c^\ast$.

Even slightly below $T_c^\ast$, $n_s(\omega=0)/m^\ast$ obtains a
significant finite value, leading to the Mei\ss{}ner effect, and there is a
considerable redistribution of weight from energies roughly above twice the
low-temperature maximum gap, $2\Delta_0$, to energies below $2\Delta_0$.
This redistribution increases with decreasing temperature. Also, a peak
develops slightly below $\Delta_0$ followed by a dip around $2\Delta_0$,
this structure being most pronounced in the underdoped case. Since
$\Delta_0$ is smaller in the overdoped regime, $n_s(\omega)/m^\ast$ changes
more rapidly for small $\omega$ in this case. It is of course not
surprising that $2\Delta_0$ is the characteristic frequency of changes in
$n_s(\omega)/m^\ast$ related to the formation of Cooper pairs.

We now turn to the question of how phase fluctuations affect the
dynamical phase stiffness $n_s(\omega)/m^\ast$. This requires a dynamical
generalization of BKT theory, which was first developed by
Ambegaokar {\it et al.}\cite{AHNS,Minn} Here, we start
from a heuristic argument for the dynamical screening of the vortex
interaction:\cite{AHNS} An applied electromagnetic field exerts
a force on the vortices mainly by inducing a superflow, which leads to a
Lorentz force on the flux carried by the vortices. On the other hand,
moving a vortex leads to dissipation in its core and thus to a finite
diffusion constant $\Dv$,\cite{BS} which impedes its motion.
If one assumes a rotating field of frequency $\omega$,
small vortex-antivortex pairs will rotate to stay aligned with the
field. Large pairs, on the other hand, will not be able to follow the
rotation and thus become ineffective for the screening.
A pair can follow the field if its component vortex and
antivortex can move a distance $2\pi r$ during one period
$T_\omega=2\pi/\omega$. During this time a vortex can move a distance of
about the diffusion length $\sqrt{\Dv T_\omega}=\sqrt{2\pi \Dv/\omega}$,
so that the critical scale for the pair size is
\be
r_\omega \equiv \sqrt{\frac{\Dv}{2\pi\omega}} .
\ee
Only vortex-antivortex pairs of size $r\lesssim r_\omega$ contribute to the
screening. Hence, we cut off the renormalization flows at this length scale.
To avoid an unphysical kink in $n_s^R(\omega)/m^\ast$ we use the smooth cutoff
$\overline{r}^2=r_\omega^2+r_0^2$.

The diffusion constant of vortices is not easy to calculate accurately. In
the absence of pinning, the theory of Bardeen and Stephen\cite{BS} yields
\be
\Dv^0 = \frac{2\pi c^2 \xi_{ab}^2 \rho_n\, k_BT}{\phi_0^2\, \tilde d} ,
\label{3.Dv0}
\ee
where $c$ is the speed of light, $\xi_{ab}\sim r_0/2$ is the
coherence length, $\rho_n$ is the normal-state resistivity,
$\phi_0=hc/2e$ is the superconducting flux quantum, and $\tilde d$
is an effective
layer thickness. In the renormalization the quantity $\Dv^0/r_0^2$ enters,
which according to Eq.~(\ref{3.Dv0}) is linear in temperature.
In the presence of a high density of weak pinning centers
the diffusion constant becomes\cite{Fish80} $\Dv=\Dv^0\,\exp(-E_p/k_BT)$,
where $E_p$ is the pinning energy. Matters are complicated by the
observation that $E_p$ depends on temperature. Rogers
{\it et al.}\cite{Roge92} find  $E_p(T) \approx E_p^0\,(1-T/T_c^\ast)$
with $E_p^0/k_B \approx 1200\:{\rm K}$ for
Bi$_2$Sr$_2$CaCu$_2$O$_{8+\delta}$.
Absorbing the constant term in the exponent into
the prefactor, the result for the diffusion constant in natural units is
\be
\frac{\Dv}{r_0^2} \approx C_v\, \frac{k_BT}{\hbar}\,
  \exp\!\left(-\frac{E_p^0}{k_BT}\right) ,
\label{3.Dv3}
\ee
where $C_v$ is a dimensionless constant.
However, such a large value of $E_p^0$ would lead to a sharp, step-like
dependence of $n_s(\omega)/m^\ast$ on temperature, in contradiction to the
smooth behavior shown in Fig.~4 of Ref.~\onlinecite{Cors}. In view of these
difficulties we treat $\Dv/r_0^2$ as a constant parameter and discuss the
dependence on $\Dv$ below.

\begin{figure}
\includegraphics[width=3.20in]{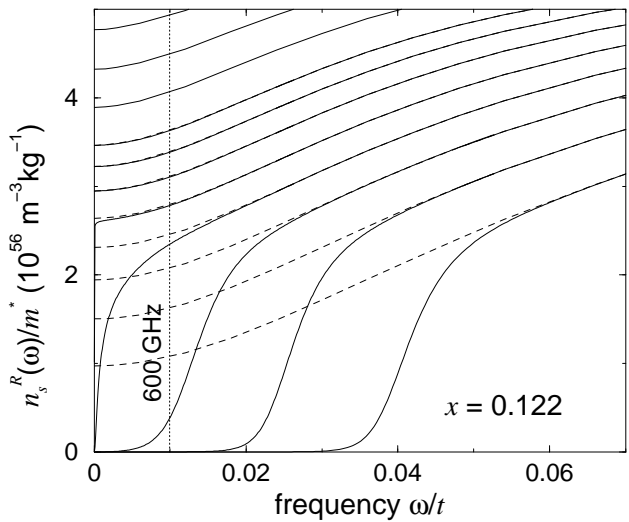}
\caption{\label{fig.nsRom}Phase stiffness $n_s^R(\omega)/m^\ast$
renormalized by vortex fluctuations for $x=0.122$ at temperatures $k_BT/t=$
0.016, 0.017, 0.018, 0.019, 0.0195, 0.02, 0.0205, 0.021, 0.0215, 0.022,
0.0225 (heavy solid lines). The vortex diffusion constant has been chosen
as $\Dv/r_0^2 = 10^{17}\,{\rm s}^{-1}$. The unrenormalized stiffness is
shown as dashed lines; these are the same data as in
Fig.~\protect\ref{fig.nsom}. Note the expanded frequency scale. The highest
frequency used by Corson {\it et al.}\cite{Cors} is indicated by the
vertical dotted line.}
\end{figure}

To find the effect of phase (vortex) fluctuations on the phase stiffness,
the recursion relations (\ref{2.Kost1}) and (\ref{2.Kost2}) are now
integrated numerically up to the cutoff
$\overline{l}=\ln(\overline{r}/r_0)$, which depends on $\Dv/r_0^2$. The
resulting renormalized phase stiffness $n_s^R(\omega)/m^\ast$ for constant
$\Dv/r_0^2 = 10^{17}\,{\rm s}^{-1}$ and $x=0.122$ is plotted in
Fig.~\ref{fig.nsRom}. Other values of $\Dv$ give similar results. Of
course, faster vortex diffusion shifts the features at given temperature to
higher frequencies. The dashed lines denote the unrenormalized stiffness,
{\it i.e.}, the same data as in Fig.~\ref{fig.nsom}, albeit on an expanded
frequency scale. The highest frequency used in Ref.~\onlinecite{Cors}
($600\,{\rm GHz}$) corresponds to $\omega/t\approx 0.01$, also indicated in
Fig.~\ref{fig.nsRom}.

For $T<T_c$ (the upper six curves) the static renormalization has been
found to be small, see Sec.~\ref{sec.2}. The renormalization at finite
$\omega$ is even weaker so that the renormalized stiffness is in practice
identical to the unrenormalized one, which has only a weak frequency
dependence for low $\omega$, in agreement with Ref.~\onlinecite{Cors}.

When $T$ is increased above $T_c$ (the lower five curves in
Fig.~\ref{fig.nsRom}), a strong renormalization of the stiffness due to
phase fluctuations sets in starting at very low frequencies. The
Mei\ss{}ner effect is thus destroyed for all $T>T_c$ by the comparatively
slow vortex diffusion. With increasing temperature the onset of
renormalization shifts to higher frequencies. At frequencies above this
onset, the vortices cannot follow the field and thus do not affect the
response, as discussed above. The onset frequencies are always much smaller
than $2\Delta_0$. The features at the energy scale $2\Delta_0$ shown in
Fig.~\ref{fig.nsom}, which are due to Cooper-pair formation, are unaffected
by phase fluctuations and show no anomality at $T_c$. They vanish only at
$T_c^\ast$.

\begin{figure}
\includegraphics[width=3.20in]{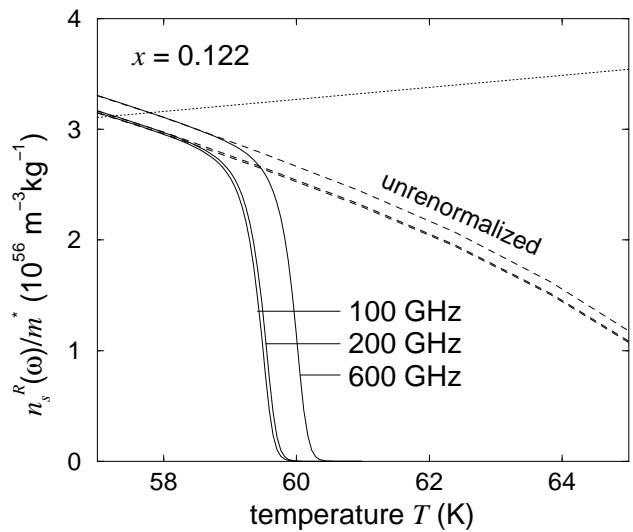}
\caption{\label{fig.nsRT}Renormalized phase stiffness
$n_s^R(\omega)/m^\ast$ for $x=0.122$ as a function of temperature for
frequencies $f=$ 100\,GHz, 200\,GHz, 600\,GHz (heavy solid lines). The
unrenormalized stiffness is shown as dashed lines. The dotted line
represents the approximate criterion Eq.~(\protect\ref{2.BKTcond}) for the
($\omega=0$) BKT transition.}
\end{figure}

Finally, in Fig.~\ref{fig.nsRT} we plot the renormalized
$n_s^R(\omega)/m^\ast$ for $x=0.122$ as a function of temperature for
various frequencies. This graph should be compared to Figs.~2 and 4 of
Ref.~\onlinecite{Cors}---note that the quantity $T_\theta$ given there is
proportional to $n_s^R/m^\ast$. We note that Corson {\it et al}.\cite{Cors}
assume a thermally activated density of free vortices, $n_f \propto
\exp(-E_f/T)$, for $T$ not too close to the BKT transition temperature, and
a temperature-independent diffusion constant.\cite{rem.Cors} Here, we
instead integrate the recursion relations (\ref{2.Kost1}) and
(\ref{2.Kost2}) explicitly up to the dynamical length scale $\overline{r}$
so that we do not have to make an assumption on $n_f$. One sees that even
at $f=600\,{\rm GHz}$ the broadened BKT transition is still much narrower
than found by Corson {\it et al.}\cite{Cors} From Eqs.~(\ref{3.Dv0}) and
(\ref{3.Dv3}) it is clear that the diffusion constant $\Dv/r_0^2$ increases
with temperature. In the presence of pinning it increases rapidly around
$k_BT \sim E_p^0$. Since a larger diffusion constant, {\it i.e.}, more
mobile vortices, leads to stronger renormalization, the transition in
Fig.~\ref{fig.nsRT} would become {\it even sharper\/} if $\Dv/r_0^2$ were
an increasing function of temperature.

Our results show that dynamical BKT theory together with Bardeen-Stephen
theory for vortex diffusion and natural assumptions on pinning does {\it
not\/} agree quantitatively with the experimental results.\cite{Cors} We
conclude that the finite size effect apparent in the experimental data is
not only due to the finite diffusion length. Another possible source is the
interlayer Josephson coupling, which leads to the apperance of the
Josephson length $\Lambda$ as an additional length scale, as discussed
above.\cite{Blt} This length scale leads to a cutoff of the recursion
relations at $l\sim \ln(\Lambda/r_0)$, which becomes small close to
$T_c^\ast$ due to the divergence of $r_0\sim \xi_{ab}$ (neglecting the
feedback of phase fluctuations on the quasiparticles). This broadens the
transition but cannot easily explain the observed frequency dependence. On
the other hand, the experimental observation that the curves for various
frequencies\cite{Cors} start to coincide where the phase stiffness agrees
with the universal jump criterion (\ref{2.BKTcond}) supports an
interpretation in terms of vortex fluctuations. We suggest that a
better description of the interplay of vortex dynamics and interlayer
coupling is required to understand the data. 

Note, the origin of the discrepancy may also lie in the FLEX results for
$n_s(\omega)/m^\ast$, which do not include all effects of
temperature-dependent scattering on the conductivity $\sigma$,\cite{Scharn}
and in the omission of the feedback of phase fluctuations on the electronic
properties. Another effect neglected here is the possible coupling to a
charge-density wave perhaps taking the form of dynamical stripes.

\section{Summary and conclusions}

In the present paper we have obtained the characteristic energy scales of
hole-doped cuprate superconductors from a theory that includes both spin
and Cooper-pair phase fluctuations. The former are described by the FLEX
approximation, whereas the latter are included by means of the
Berezinskii-Kosterlitz-Thouless (BKT) theory, taking the FLEX results as
input. Phase fluctuations mainly take the form of {\it vortex\/}
fluctuations, since Gaussian phase flucuations have a large energy gap.
Vortices lead to the renormalization of the phase stiffness
$n_s(\omega)/m^\ast$ to $n_s^R(\omega)/m^\ast$. The stiffness at $T\to 0$
shows a maximum at a doping level of $x\approx 0.2$, in good agreement with
experiments.\cite{Bern} At the transition temperature $T_c$ the
renormalized static phase stiffness $n_s^R(\omega=0)/m^\ast$ vanishes,
leading to the disappearance of the Mei\ss{}ner effect. The ideal
conductivity is also destroyed by free vortices. $T_c$ is significantly
reduced compared to the transition temperature $T_c^\ast$ that would result
from spin fluctuations alone. The $T_c$ determined from spin {\it and\/}
phase (vortex) fluctuations is in much better agreement with experiments
in the underdoped regime and shows a maximum at optimum doping. Still,
our approach does not explain the full reduction of $T_c$. We
believe that a further reduction of $T_c$ results from (a) the breaking of
Cooper pairs by scattering with phase fluctuations and (b) other
instabilities that reduce the density of states in the normal state, for
example a charge-density wave. Since the latter effect also suppresses
$n_s/m^\ast$, phase fluctuations can become even more effective and reduce
$T_c$ further. It would be desirable to include the pair-breaking effect of
phase fluctuations and the possible formation of a charge-density wave on
the same microscopic level as the spin fluctuations.\cite{FS}

For $T_c<T<T_c^\ast$, where phase-coherent superconductivity is absent,
phase fluctuations lead to a strong renormalization of $n_s/m^\ast$ at
frequencies much smaller than $2\Delta_0$.  Our results show the same
trends as found in conductivity measurements.\cite{Cors} However, a
three-dimensional description of vortex dynamics might be required to
obtain a more quantitative agreement. Local formation of Cooper pairs still
takes place in this regime. This leads to a strong pseudogap of the same
magnitude $\Delta_0$ and symmetry as the superconducting gap below $T_c$.
We also find a frequency dependence of $n_S^R(\omega)/m^\ast$ at higher
frequencies, $\omega\gtrsim \Delta_0$, that is very similar to the
superconducting phase. These features vanish only around $T_c^\ast$.
Finally, for $T_c^\ast<T<T^\ast$ there is a weak suppression in the density
of states at the Fermi energy. Our results reproduce several of the main
features common to all hole-doped cuprate superconductors. We conclude that
the exchange of spin fluctuations, modified by strong superconducting phase
(vortex) fluctuations in the underdoped regime, is the main mechanism of
superconductivity in cuprates.

\acknowledgements

We would like to thank I. Eremin, H. Kleinert, F. Sch\"afer, and D. Straub
for helpful discussions.


\begin{thebibliography}{99}
\bibitem{rem.gauss}In 2D Gaussian fluctuations destroy long-range order.
However, the transition between quasi-long-range and short-range order
still takes place at the mean-field $T_c$ if only Gaussian fluctuations
are taken into account.
\bibitem{rem.fieldth}This derivation\protect\cite{KD,SBL,FS} starts from an
attractive pairing interaction, which is decoupled using a
Hubbard-Stratonovich transformation. The decoupling field is the complex
superconducting gap $\Delta$. A gauge transformation of the fermions then
removes the phase from $\Delta$ and couples it directly to the fermions.
After the fermions are integrated out, one obtains the effective action of
the phase.
\bibitem{rem.EK}The quantity $V_0$ in Ref.~\onlinecite{EK95} is related to
$K$ by $K=\beta V_0$.
\bibitem{rem.y}One usually writes $y=N_0 e^{-\beta E_c}$, where $N_0$ is a
numerical factor of order unity. However, the results depend on $N_0$ only
very weakly, since the renormalization of $n_s$ below $T_c$ is small.
\bibitem{rem.Cors}Note that the quantity shown in the inset in Fig.~3
of Ref.~\protect\onlinecite{Cors} is $1/(\Theta\Omega)$ instead of
$\Theta\Omega$ and that in the caption one should have
$\Omega=(\Omega_0/\Theta)\,\exp(2C/\Theta)$.
\end{thebibliography}
\end{document}